\documentstyle[11pt,moriond,epsfig]{article}

\def\be{\begin{equation}}
\def\ee{\end{equation}}
\def\bea{\begin{eqnarray}}
\def\eea{\end{eqnarray}}

\begin{document}
\vspace*{4cm}
\title{QCD AT LEP\,2 AND WW FINAL STATE INTERACTIONS \footnote{to be
published in the proceedings of the $34^{th}$ Rencontres de Moriond, Les
  Arcs 1800, France, 13-20 March 1999}}

\author{ M.~HAPKE }

\address{Queen Mary and Westfield College, University of London,\\
London E1 4NS, UK}

\maketitle\abstracts{A short overview of the QCD program at LEP~2 is given.
Studies of final state interactions in  ${\rm e}^+{\rm e}^- \rightarrow
{\rm W}^+{\rm W}^-\rightarrow 
{\rm q}\bar{\rm q}^\prime{\rm q}\bar{\rm q}^\prime$ decays are discussed.
}

\newpage
\section{Introduction}
After the studies of ${\rm e}^+{\rm e}^-$ collisions at a centre-of-mass 
energy of $\sqrt{s} = M_{\rm Z^0}$ at LEP~1, the LEP~2 program has so far
allowed the study
of ${\rm e}^+{\rm e}^-$ collisions at $\sqrt{s} =$ 133, 161, 172, 183 and 189
GeV. While at LEP~1, the collected data samples of typically 4 million Z$_0$
decays per experiment allow high precision QCD tests, one of the aims at LEP~2 is the study of the
energy dependence of QCD quantities using non-radiative 
${\rm e}^+{\rm e}^- \rightarrow  {\rm q}\bar{\rm q}$ events. Typical data samples
consist of several hundred events at lower LEP~2 energies going up to 
roughly 3000 events at $\sqrt{s} =$ 189 GeV, the centre-of-mass energy for 
data collected in 1998.

Effects of the strong interaction are not only studied in 
${\rm e}^+{\rm e}^- \rightarrow  {\rm q}\bar{\rm q}$ events, but also
in ${\rm e}^+{\rm e}^- \rightarrow  {\rm W^+}{\rm W^-}$ events
where both W bosons decay hadronically, ${\rm W }\rightarrow 
{\rm q}\bar{\rm q}^\prime$. The decay vertices of the two W bosons are on average
less than 0.1 fm apart, which is small compared with the typical hadronic length
scale of 1 fm. It is therefore possible that the decays of the two Ws are not
independent but are affected by interactions
between quarks from the decays of the two different W bosons via
{\sl colour reconnection} 
or by interactions between the hadrons produced in the hadronisation of the two 
W bosons via {\sl Bose-Einstein correlations}.
The study of these effects is interesting in its own right and is
of special importance for the measurement of the W mass in fully hadronic W
decays where the uncertainty due to possible WW final state interactions is
currently the largest contribution to the systematic uncertainty \cite{riu}. 
The number of selected WW events per experiment is typically 400 for
$\sqrt{s} =$ 183 GeV and 1500 for $\sqrt{s} =$ 189 GeV.
 
\section{QCD at LEP\,2}

\begin{figure}[htb]
\setlength{\unitlength}{1mm}
\begin{picture}(140,70)
\setlength{\unitlength}{1mm}
\put(0,2){\resizebox{70mm}{!}
{\includegraphics*{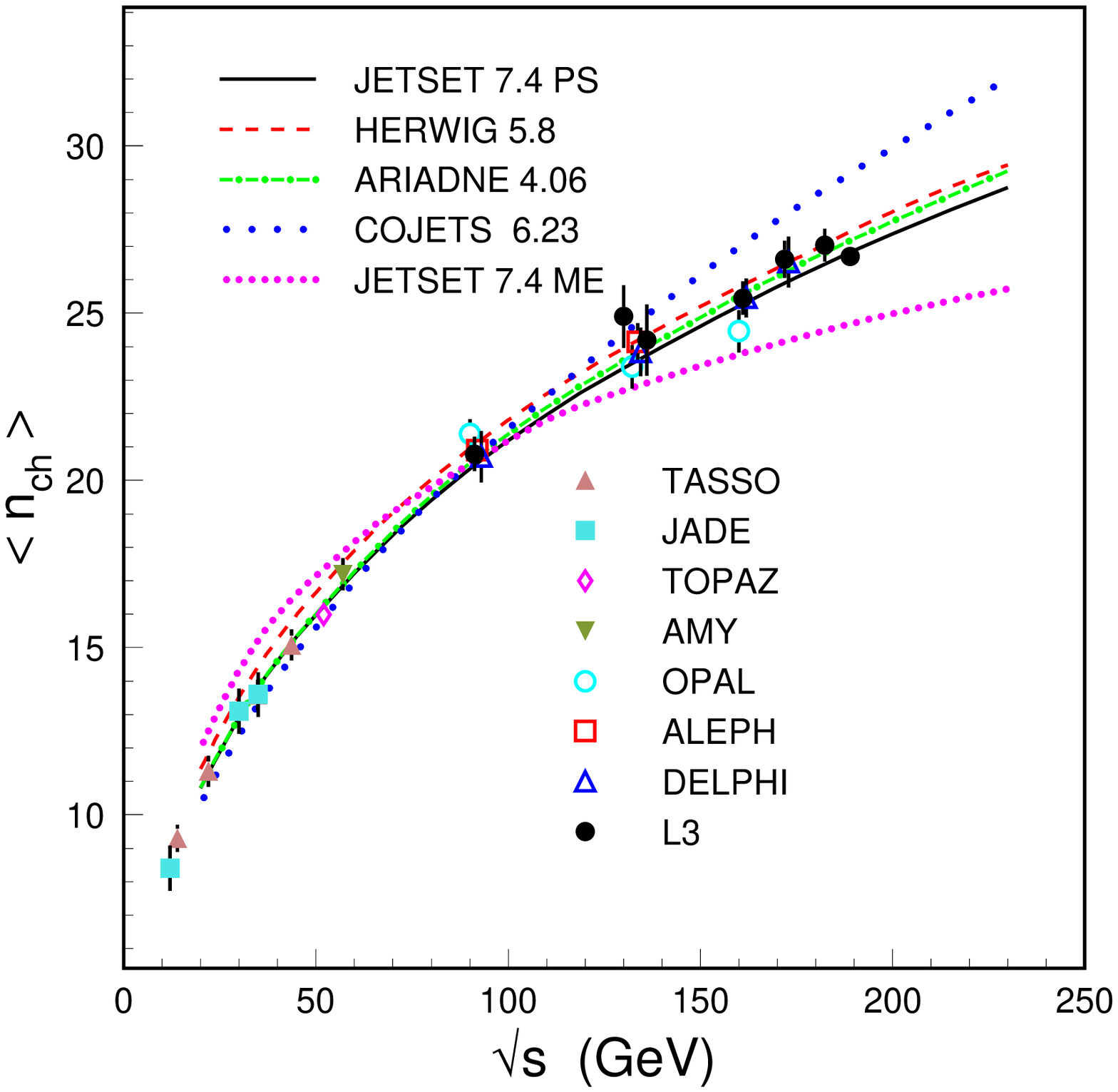}}}
\put(62,-61){\resizebox{120mm}{!}
{\includegraphics*{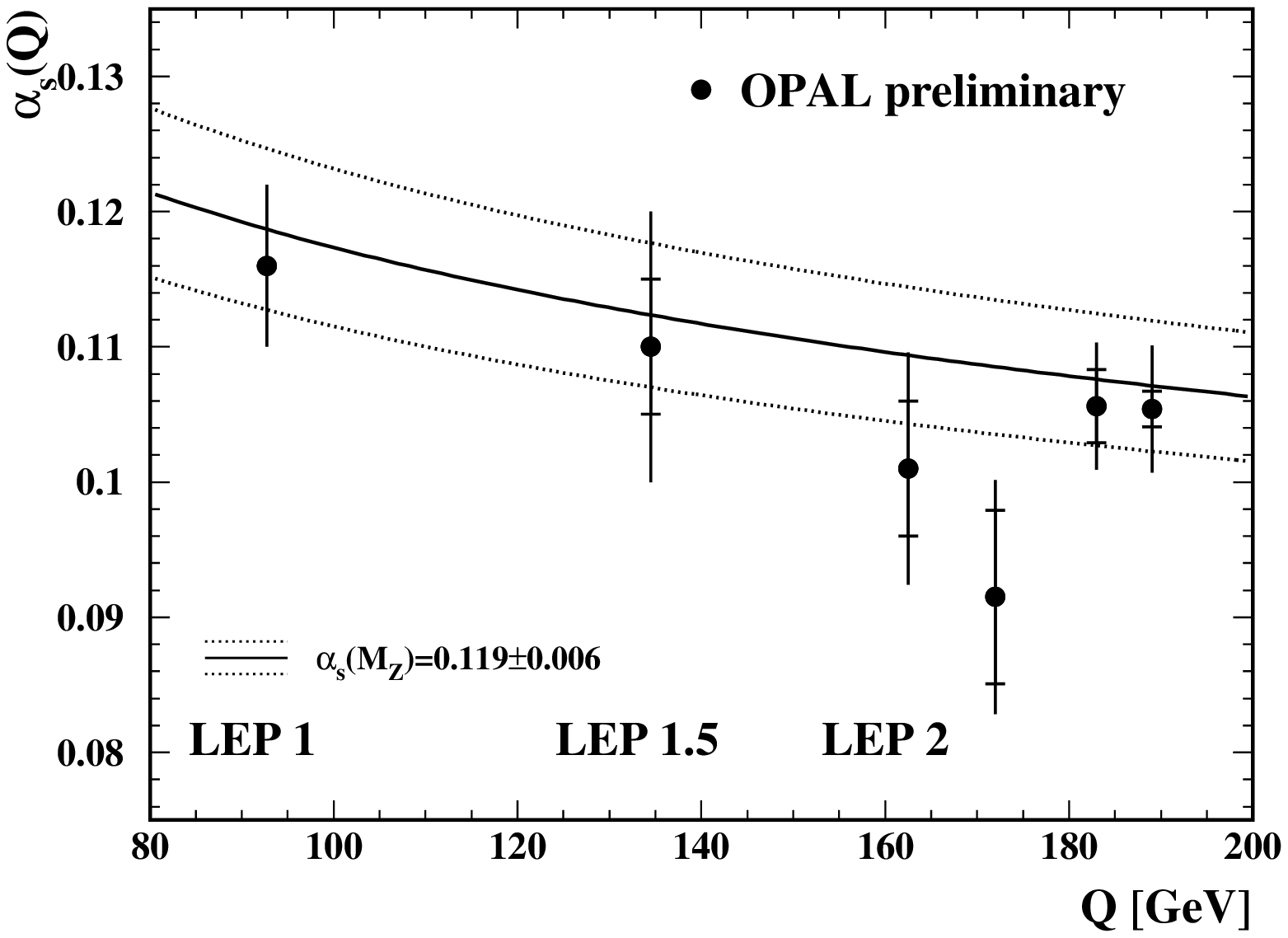}}}
\end{picture}
\caption{
On the left: published measurements of the mean charged multiplicity at
different centre-of-mass energies, together with preliminary L3 results and 
model
predictions $^3$. On the right: preliminary OPAL results for $\alpha_s$ 
at different centre-of-mass energies $^4$. The error bars show the
total errors, the extent of the statistical errors is indicated by small 
horizontal bars.
\label{fig:mult}
}
\end{figure}

General features of hadronic events such as event shapes, jet rates or charged
particle momentum spectra are studied by all LEP collaborations at all
centre-of-mass energies. A very basic quantity is the 
mean charged particle
multiplicity, for which results at the highest centre-of-mass energy of $\sqrt{s} =$
189 GeV are already available from three collaborations \cite{a189,l189,o189} as
summarised in table
\ref{tab:mult}. The energy evolution of this quantity
is shown in fig.~\ref{fig:mult} and illustrates the remarkably good agreement
between measurements and the predictions of parton
shower models like JETSET, HERWIG and ARIADNE which include QCD coherence
effects and which were tuned to describe the data at LEP~1. 
Parton shower models with no QCD coherence effects like
COJETS or matrix element models as implemented in JETSET cannot explain the
energy dependence. Good agreement between data and JETSET, HERWIG and ARIADNE
predictions are also observed for various charged particle momentum
distributions, jet rates and for event shapes like thrust.  

\begin{table}[htb]
\caption{
Mean charged multiplicity 
$\langle n_{ch} \rangle $ and $\alpha_s$ as
determined from fits to the event shapes at a centre-of-mass energy of 
$\sqrt{s} =$189 GeV. All numbers are preliminary. 
}
\label{tab:mult}
\vspace{0.4cm}
\begin{center}
\begin{tabular}{|c|c|c|}
\hline
Experiment & $\langle n_{ch} \rangle $ & $\alpha_s$ \\
\hline
ALEPH & $27.37 \pm 0.20 ({\rm stat})  \pm 0.25 ({\rm syst})$
      & $0.1119\pm0.0015({\rm stat})\pm 0.0011({\rm exp})\pm0.0030({\rm theo})$ 
      \\
L3 & $26.73 \pm 0.13 ({\rm stat})  \pm 0.18 ({\rm syst})$
   & $0.1082\pm 0.0028({\rm exp}) \pm 0.0052({\rm theo})$
   \\
OPAL & $26.83 \pm 0.16 ({\rm stat})  \pm 0.30 ({\rm syst})$
      & $0.107 \pm 0.001({\rm stat})  \pm 0.005({\rm syst})$ 
      \\
 \hline
\end{tabular}
\end{center}
\end{table}

Event shapes are not only used to compare data and model predictions. Some 
event shape variables 
are predicted by ${\cal O} (\alpha_s^2)$ calculations combined 
with predictions of the next to leading log approximation (NLLA) of QCD.
A fit of these  ${\cal O} (\alpha_s^2)+$NLLA predictions to the measured
event shapes allows the extraction of $\alpha_s$. Up to six different event 
shapes
have been used, and the combined results for $\alpha_s$ at $\sqrt{s} =$  189 GeV
are presented in table \ref{tab:mult}. In fig.~\ref{fig:mult}, the energy
dependence of these results is shown to be consistent with the expected energy
evolution of $\alpha_s$. It is interesting to note that the uncertainty of the
results at the highest energy points is no longer dominated by the statistical
uncertainty and the total uncertainty is very similar to the total uncertainty
of the $\alpha_s$ measurement at $\sqrt{s} = M_{Z^0}$.

The energy evolution of the mean value of some event shape distributions can be
used to determine $\alpha_s$ using the power law ansatz of Dokshitzer and Webber
\cite{DW}. In this ansatz non-perturbative effects depend only on the energy
scale, $\alpha_s$ and a new universal parameter $\alpha_0$. ALEPH \cite{apow}
and DELPHI \cite{dpow} fitted the results for the main values of event shape
variables  to determine $\alpha_s$ and $\alpha_0$. 
Results for $\alpha_s$ are consistent with the world average and the numbers
measured for $\alpha_0$ are consistent with a common value for all 
studied event shapes and with the values obtained from a study of re-analysed 
JADE data \cite{jpow} and from lepton-proton scattering at H1 \cite{hpow}.

\section{WW Final State Interactions}

\subsection{Colour Reconnection}

The decay products of the two W decays in 
${\rm e}^+{\rm e}^- \rightarrow  {\rm W^+}{\rm W^-}$ events may have a
significant space-time overlap as the separation of their decay vertices is 
small compared to characteristic hadronic distance scales. In the fully 
hadronic channel this may lead to new types of final state interactions.
Colour reconnection is the general name applied to the case where such final 
state interactions lead to colour flow between the decay products of the two W
bosons. At present there is general consensus  that observable effects of such
interactions during the perturbative phase are expected to be small \cite{ocr}.
In contrast, significant
interference in the hadronisation process is considered to be a real
possibility. With the current knowledge of non-perturbative QCD, such
interference can be estimated only in the context of specific models.

Studies of reconnection phenomena implemented with the ARIADNE, PYTHIA and HERWIG
models show that these effects might lead to a reduction of the mean charged
multiplicity by up to approximately 3$\%$ when compared to a situation 
in the absence of colour reconnection effects \cite{ocr,acr,dcr,lcr}.   
Studies using the Ellis-Geiger model predicted a reduction of the order of 10
$\%$ \cite{EG}. However, the Monte Carlo program VNI in which the Ellis-Geiger
model is implemented does not describe the data \cite{ocr,acr}. Therefore, the
present implementation of the VNI generator is not used to study effects of
colour reconnection.

Some earlier estimates of the sensitivity to colour reconnection have been made
within the context of given models, comparing ``reconnection'' to ``no
reconnection'' scenarios for  ${\rm W}^+{\rm W}^- \rightarrow 
{\rm q}\bar{\rm q}^\prime{\rm q}\bar{\rm q}^\prime$ events. In general, both the
size and sign of any changes are dependent upon the model considered. At the
expense of a reduction in statistical sensitivity, the dependence on the
modelling of single hadronic W decays can be avoided by comparing directly with 
the properties of the hadronic part of ${\rm W}^+{\rm W}^- \rightarrow 
{\rm q}\bar{\rm q}^\prime\ell\bar{\nu}^\prime$ events.
All four LEP experiments compared the mean charged particle multiplicity in
fully hadronic events $\langle n_{\rm ch}^{4\rm q} \rangle$ and twice the
charged particle multiplicity in semileptonic events 
$2\langle n_{\rm ch}^{\rm qq\ell\bar{\nu}} \rangle$ where the charged
particles associated with the leptonically decaying W are excluded. 
Results of these comparisons
are shown in table \ref{tab:cr}. Most results show no significant difference
between $\langle n_{\rm ch}^{4\rm q} \rangle$ and 
$2\langle n_{\rm ch}^{\rm qq\ell\nu} \rangle$, however, the uncertainty on the results
are of the order of the effects predicted by the colour reconnection models
implemented in PYTHIA, ARIADNE and HERWIG and only extreme models can be
excluded so far. However, it is interesting to note that the DELPHI results
for the ratio ${\langle n_{\rm ch}^{4\rm q} \rangle} /
{2\langle n_{\rm ch}^{\rm qq \ell \bar{\nu}} \rangle}$ differ from unity
by one standard deviation for $\sqrt{s} =$ 189
GeV and two standard deviations for $\sqrt{s} =$183 GeV. DELPHI have also 
measured this ratio for soft particles with momenta between 
0.1 and 1 GeV where the effect is expected to be more pronounced. 
Preliminary results for this  ratio in this
restricted momentum range are $0.926 \pm 0.025 \pm 0.023$ for the $\sqrt{s}
=$ 183 GeV data and $0.966 \pm 0.017 \pm 0.027$ for $\sqrt{s} =$ 189 GeV. 
\begin{table}[htb]
\caption{
Difference and ratio of the mean charged particle multiplicities for fully
hadronic and semileptonic WW events. All results but those from OPAL are
preliminary.
}
\label{tab:cr}
\vspace{0.4cm}
\begin{center}
\begin{tabular}{|c|c|c|c|}
\hline
Experiment & $\sqrt{s}$ & Studied Quantity & Result  \\
\hline
\begin{minipage}{2cm}{ALEPH \\ ALEPH \\ L3 \\ OPAL}\end{minipage} 
&
\begin{minipage}{2cm}{183 GeV \\ 189 GeV \\ 183 GeV \\ 183 GeV}\end{minipage}
&
$\langle n_{\rm ch}^{4\rm q} \rangle - 
2\langle n_{\rm ch}^{\rm qq\ell\bar{\nu}} \rangle$
&
\begin{minipage}{4cm}{\begin{tabular}{c @{$\pm$} c @{$\pm$} c@{ } c }
                      1.31 & 0.74 & 0.37 \\ 
                      0.47 & 0.44 & 0.26 \\ 
                      -1.0 & 0.8  & 0.5  \\ 
                      0.7 & 0.8  & 0.6  \\ 
		      \end{tabular}
		      }\end{minipage}
\\
\hline
\begin{minipage}{2cm}{DELPHI \\ DELPHI}\end{minipage} 
&
\begin{minipage}{2cm}{183 GeV \\ 189 GeV}\end{minipage}
&
${\langle n_{\rm ch}^{4\rm q} \rangle} /
      {2\langle n_{\rm ch}^{\rm qq \ell \bar{\nu}} \rangle}$
&
\begin{minipage}{4cm}{\begin{tabular}{c @{$\pm$} c @{$\pm$} c@{ } c }
                      0.941 & 0.025 & 0.023 \\ 
                      0.977 & 0.017 & 0.027 \\ 
		      \end{tabular}
		      }\end{minipage}
\\
 \hline
\end{tabular}
\end{center}
\end{table}

Not only has the mean value of the charged particle multiplicity been studied,
but also its dispersion. Additionally the rapidity and $p_t$ distributions 
of charged
tracks in fully hadronic events have been compared to those distributions in
semileptonic events. DELPHI also studied the mean multiplicity 
of heavy hadrons in the low momentum region \cite{dcr2},
where the effects of
colour reconnection may be further enhanced.
However, none of these  results show clear evidence for colour
reconnection effects.

\subsection{Bose-Einstein Correlations}

Bose-Einstein correlations (BEC) between  pairs of identical bosons, mainly the
$\pi^\pm\pi^\pm$ system, have been extensively studied in a large variety of interactions and
over a wide range of energies. Formally, they can be expressed by
the correlation function
\begin{equation}
\label{eqn:1}
R(p_1,p_2) = \frac{\rho(p_1,p_2)}{\rho_0(p_1,p_2)}
\end{equation}
where $p_1,p_2$ are the four-momenta of the bosons, $\rho$ is the two-particle
probability density and $\rho_0$ is the two-particle probability density that would occur
in the absence of Bose-Einstein correlations. For the study of BEC in WW events
ALEPH \cite{abec}, L3 \cite{lbec} and OPAL \cite{obec} used a reference
distribution $\rho_0$ based on the observed two-particle probability density of
particle pairs with opposite sign, while DELPHI \cite{dbec} used the
prediction for $\rho_0$ from Monte Carlo models that do not include 
Bose-Einstein effects.
The correlation is large for small four-momentum differences, $
Q = \sqrt{-(p_1-p_2)^2}$,
so that often Bose-Einstein correlations are parametrised in terms of
this one-dimensional
distance measure. A common parametrisation is
\begin{equation}
\label{eqn:3}
R(Q) = 1 + \lambda e^{-Q^2r^2}
\end{equation}
where $r$ estimates the size of the two-boson emitter which is taken to be of Gaussian
shape and $\lambda$ is a measure of the strength of the Bose-Einstein effect.

\begin{figure}[htb]
%\rule{5cm}{0.2mm}\hfill\rule{5cm}{0.2mm}
%\vskip 2.5cm
%\\
%\rule{5cm}{0.2mm}\hfill\rule{5cm}{0.2mm}
%\psfig{figure=mult.eps,height=1.5in}
\setlength{\unitlength}{1mm}
\begin{picture}(150,103)
\setlength{\unitlength}{1mm}
\put(-4,0){\resizebox{100mm}{!}
{\includegraphics*{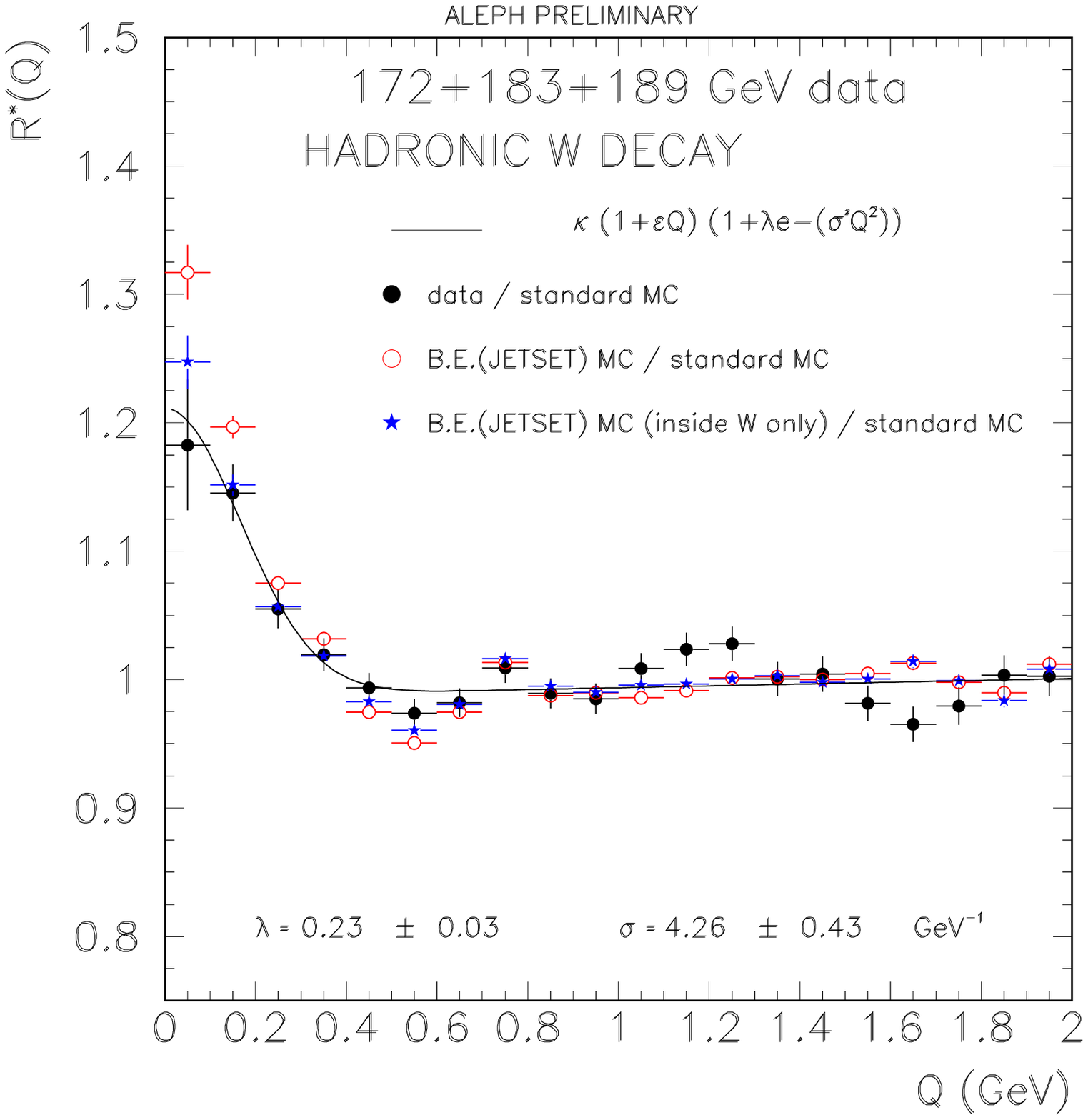}}}
\put(73,-42){\resizebox{120mm}{!}
{\includegraphics*{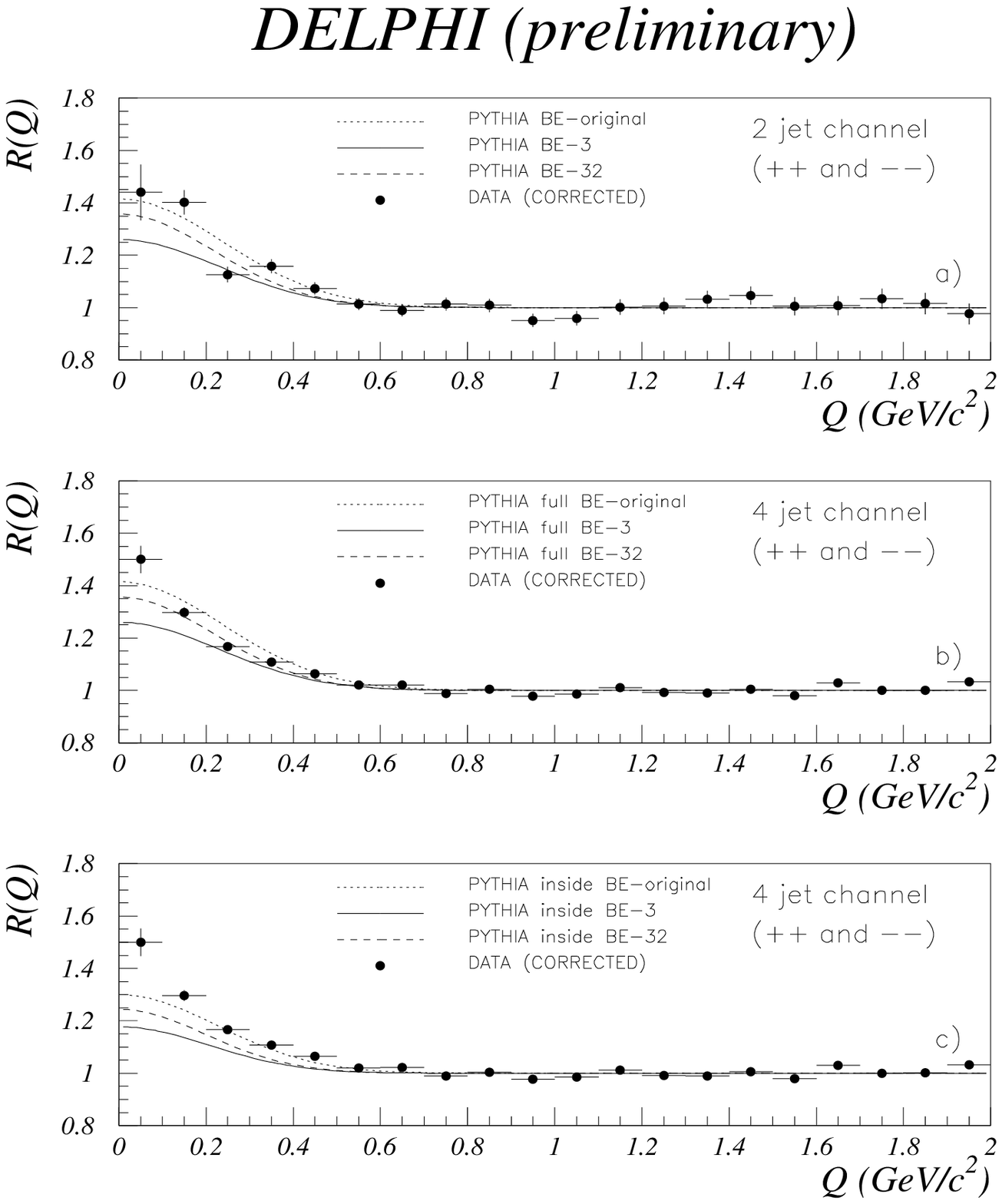}}}
\end{picture}
\caption{Results for the correlation function in fully hadronic events
together with MC predictions from ALEPH (left) and correlation functions 
for semileptonic and fully hadronic WW events together with MC prediction
from DELPHI (right).
\label{fig:bec}}
\end{figure}

The current studies of BEC in W events are motivated by the question of whether
BEC for pions from different W bosons exist or not. An attempt to answer this
question is to compare the measured correlation function for fully hadronic WW
events with MC models that either include BEC among bosons from different Ws or 
include BEC only inside each of the two W systems.
ALEPH as well as DELPHI have already analysed the $\sqrt{s} =$ 189 GeV data
sample and the results are shown in fig.~\ref{fig:bec}. However, while ALEPH
conclude from the comparison of data with carefully tuned JETSET model
predictions that BEC between decay products of different W is
disfavoured at a 2.7$\sigma$ level, DELPHI now observe --- in contrast 
to their previous results \cite{dbec3} --- an indication of the
presence of BEC between particles from different W bosons.  

All four LEP collaborations attempted to measure directly the strength of the
BEC between  particles from different W bosons. In an analysis of  the
$\sqrt{s}=172$ data, DELPHI used the two particle densities in 
semileptonic WW events to subtract the contribution of particle pairs coming
from the same W boson from the observed two-particle densities in fully hadronic
events. The correlation strength of the remaining distribution is
$\lambda^{\rm diff\ W} = -0.20 \pm 0.22 \pm 0.08$. ALEPH used a similar method and
obtained for data with $\sqrt{s} =$ 172 and 183 GeV 
a value of $\lambda^{\rm diff\ W} = 0.15 \pm 0.18 (stat)$ with a
negligible systematic error. L3 studied fully hadronic events at $\sqrt{s} =$
183 GeV and assigned
jet pairs to the different W bosons, allowing a direct determination of 
the correlation strength of  $\lambda^{\rm diff\ W} = 0.75 \pm 1.80$. OPAL
performed a simultaneous fit to the correlation functions of 
fully hadronic WW events, semileptonic WW events and 
${\rm e}^+{\rm e}^- \rightarrow  {\rm q}\bar{\rm q}$ events. Based on
the data collected at $\sqrt{s} =$ 172 and 183 GeV, the value
$\lambda^{\rm diff\ W} = 0.22 \pm 0.53 \pm 0.14$ was obtained.
 While all these published (DELPHI, OPAL) and
preliminary (ALEPH, L3) 
results for $\lambda^{\rm diff\ W}$ are consistent with zero, i.e.~compatible
with the
absence of BEC between particles from different W bosons, the most recent
preliminary DELPHI
results indicate evidence for BEC between particles from different Ws.
The study is based on the data samples at $\sqrt{s} =$ 183 and 189 GeV and uses
a reference sample from  
 ``mixed events'' which were constructed out of two semileptonic WW events.
The result is $\lambda^{\rm diff\ W} = 
(0.073 \pm 0.025 \pm 0.018)\times f$ where $f$ is the fraction of like-sign
particle pairs from different Ws in the fully hadronic WW channel at low Q, and
was found to be 0.201 when calculated using PYTHIA MC events.

\section{Conclusions}

The QCD studies of the four LEP experiments in 
${\rm e}^+{\rm e}^-$ collisions at a centre-of-mass 
energy of up to 189 GeV have resulted in measurements showing good agreement
with model predictions.
While extreme models for colour reconnection effects in fully 
hadronic ${\rm e}^+{\rm e}^- \rightarrow {\rm W}^+{\rm W}^-$ events can 
be excluded by the data there is as yet no conclusive answer to the 
question of whether or 
not there are Bose-Einstein correlations between particles from different W bosons.
 
%\section*{Acknowledgements}

%Its a pleasure to thank my colleges from the QCD and W-physics working group of
%the four LEP experiments for their help in preparing this contribution and 
%the Royal Society for the financial support.
%LEP and Royal Society

\end{document}